# A STEP FORWARD TO COMPONENT-BASED SOFTWARE COST ESTIMATION IN OBJECT-ORIENTED ENVIRONMENT

M. Nadeem, M. R. Asim and M. R. J. Qureshi

Department of Computer Science, COMSATS Institute of Information Technology Lahore, Defence Road off Raiwind Road, Lahore Pakistan.
Email: {nahmed.mrasim.rjamil}@ciitlahore.edu.pk, nahmed_ciitlahore@yahoo.com

**ABSTRACT:** Software cost estimation (SCE) of a project is pivotal to the acceptance or rejection of the development of software project. Various SCE techniques have been in practice with their own strengths and limitations. The latest of these is object-oriented one. Currently object-oriented approach for SCE is based on Line of Code (LOC), function points, functions and classes etc. Relatively less attention has been paid to the SCE in component-based software engineering (CBSE). So there is a pressing need to search parameters/variables that have a vital role for the SCE using CBSE which is taken up in this paper. This paper further looks at level of significance of all the parameters/variables thus searched. The time is being used as an independent variable because time is a parameter which is almost, all previous in one. Therefore this approach may be in a way an alternate of all previous approaches. Infact the underlying research ultimately may lead towards SCE of complex systems, using CBSE, in a scientific, systematic and comprehensive way.

## INTRODUCTION

SCE is an important pre-development activity before starting actual development of software system. This can be guessed by the quote of Alfred M. Pietasanta in 1968, "Anyone who expects a quick and easy solution to the multifaceted problem of resource estimation is going to be disappointed" (Laird, 2006). Thirty years later, (Laird, 2006) and his colleagues observed that "Despite the large number of cost factors collected and the rigorous data collection, a lot of uncertainty in the estimated (entity) can be observed". The main strength of CBSE is reusability (Qureshi and Hussain, 2008). Once a component has been developed, tested and validated by the client, this component can be reused in the all future applications depending upon the requirements and functionalities. Reusability of a component will result in reduction of cost, time and effort.

CBSE is a process that emphasizes the design and construction of component-based systems using reusable software components (Roger, 2000; Qureshi and Hussain, 2008; Qureshi, 2006). The component is an independent part of a system having complete functionality. The components can be divided into two groups namely business and infrastructure components.

The two new terms namely componentization and product-line software engineering was introduced by (Mcilroy *et al.*, 1969), that's why, he is known as a pioneer in the discipline of CBSE.

There are many SCE methods (Boehm and Valerdi, 2008; Nasir, 2006; Musilek *et al.*, 2002) namely, LOC (Verner and Tate, 1992); Source Lines of Codes (SLOC) (Albrecht and Gaftney, 1983), Software Life Cycle Management (SLIM) (Putnam, 1978; Putnam and Myers, 1992), Object-oriented Metrics, Function-Point (FP) (Albrecht, 1979; Symons, 1988), COnstructive COst MOdel-I (COCOMO-I) (Boehm, 1981) and COnstructive COst MOdel-II (COCOMO-II) (Boehm, 1999). The best known SCE techniques are COCOMO-I and COCOMO-II. Function Point, was first introduced by Albrecht in 1979 (Albrecht, 1979). COCOMO-I and COCOMO-II (Boehm and Valerdi, 2008; Mahmood and Lai, 2009) were introduced by Barry Bohem. SCE techniques mainly depend upon functionality of system to be developed.

The effort (Keung *et al.*, 2004) has been done to meet the challenge of introducing a new SCE technology for a small-scale software organization. The authors (Agarwal *et al.*, 2001) discussed different techniques to measure software cost to provide a comparative analysis.

The primary aspiration of our research is SCE using CBSE. It may be mentioned that the measure of pre-development software cost depends upon the development time as an independent variable. As such this research is an attempt to switch to SCE using CBSE on Algorithmic Cost Methods (Models) (Attarzadeh and Ow, 2010) and neither on Expert Judgment (Attarzadeh and Ow, 2010) nor Machine Learning Methods (Attarzadeh and Ow, 2010). It is so because SCE of most of the software projects is done by Algorithmic Methods. Survey, case study, architectural study and experimental study are common research methodologies used to validate a research. Survey using questionnaire as a data collection instrument is selected to evaluate this research due to type/nature of research problem.

The very next section II illustrates the related work. In section III; problem statement is defined; the proposed solution of the problem is provided in section IV; results of the research is described in section V; Validity of research given in section VI; conclusion is presented in section VII; in the end the limitations and future work are discussed in section VIII.

**Related work:** The authors (Boehm and Valerdi, 2008) proposed evaluation criteria for the validity of the process models and they provided effective results. This article also explained the strengths and weaknesses of various cost estimation techniques for the period of 1965 to 2005 (40 years). Cocomo-II (Boehm, 1999) was an excellent model up to 2005 but it did not enfold the new requirement and development styles for the reuseness or estimation of cost. Cocomo-II directed the software experts to create and designed new models such as the Chinese government version of Cocomo (Cogomo) and the Constructive Commercial-off-the-Shelf Cost Model (Cocots) etc. Different future challenges were discussed for the invention of new model/methods and tools.

The author discussed different SCE techniques and highlighted various hot areas and challenges of research in the field of software cost estimation. The authors (Zaid *et al.*, 2008) emphasized that there should be a need to research more in this field to open the new horizons for novice researchers.

The author (Nasir, 2006) discussed the strengths and weaknesses of various software estimation techniques to provide the basis for the exactness of software cost estimation. Basic Project Estimation Process also presented in a wonderful style. This paper clearly elaborated the different types of models those were derived from COCOMO (I&II).

The author (Laird, 2006) described the limitations of estimations. She discussed different uncertainties of estimation and also explained that where we are and do we care how well we estimate. It was mentioned that which stage is best for best estimate. Three golden rules of estimation were also given by her with a clear cut definition and elaboration.

The sensitivity of Cocomo-II SCE model was found with the help of three methods namely Mathematical Analysis of the Estimating Equation, Monte Carlo Simulation and Error Propagation. The main focus of the authors was to prove the sensitivity of effort "E" to effort Multiplier "EM", Sensitivity of effort "E" to size "S" and Sensitivity of effort "E" to exponent scale factors "$W_i$". The authors (Musilek *et al.*, 2002) have proposed fuzzy set technique to handle the uncertainty for the software cost estimation.

The comparison and review of 12 object-oriented software metrics had proposed in 90s by Chidamber, Kemerer and Li (Koh *et al.*, 2008). According to this paper, it was broadly accepted that sizing of different types of software deliverable items was one of the most central feature of software cost estimation. The presenters also believed that by categorizing the services provided for one classes might be more practical than CTM (Coupling through message passing).

An extension of UML (Unified Modeling Language) to RE-UML (Requirements Engineering - UML) is presented by the author (Mahmood and Lai, 2009). RE-UML enabled a system analyst to find accurate candidate components those fulfilled the stakeholders' requirements. They developed a Requirements Analysis and Assessment Process (RAAP) framework that guided stakeholders for balancing user's expectations against available components. One of the main reasons of this research was the lack of Component-Based System (CBS) development phases in the UML particularly requirements analysis and component selection. According to them, RE-UML removed the need for a system analyst to learn the new notations to model CBS requirements and component selection process.

The author (Aoyama, 1998) discussed that how component-based software engineering helped out to change the way of software development. Detailed comparison of conventional reuse and CBSE (component reusability) and Development methodologies of both conventional and CBSE had elaborated with the help of diagrams/models. He also gave his vision to the new age of software development with CBSE by re-thinking different aspects of software development.

A guideline for the selection of components from the available candidate components with the help of graph-based model for component-based software development was proposed (Sedigh-Ali and Ghafoor, 2005). Optimization of components was also discussed to make the system more efficient. The core objective of their research was on cost and quality of the software system to be developed and establishing the uniform effective relation between the two.

The authors (Oses *et al.*, 2004) explained the critical issues and different architectures of component-based software simulations and also mentioned the solutions of different component problems. They suggested that there should be a proper organization who established standards to ensure trust, to support component documentation and ways to share the advantages of component-based modeling and simulation between component developers and model developers.

Reusability of components in Component Based Development (CBD) is illustrated in (Qureshi and Hussain, 2008). The author also discussed and compared different architectures of CBD. It may be mentioned that a detail explanation of advantages and disadvantages of CBD elaborated very nicely. The authors in this paper (Qureshi, 2006) presented the comparison of component-based development (CBD) with other traditional software development practices. This paper evaluated object oriented process model and author emphasized to get full benefits of reuse. The role of repository for CBD has also been discussed.

The authors of (Succi and Baruchelli, 1997) highlighted the importance of standardization of components for the software reusability. The discussion of this research paper was to find how total development cost of a software system affected on the basis of component-based software engineering. The main two factors those were affecting the standardization cost of a component have been explained. According to them, the cost of the standardization of component(s) must be included during the cost-benefit analysis of a software system.

The author (Gill, 2003) highlighted the pertinent issues of software reusability for component based development on the basis of CBSE, considered the important issues of software reusability and high level reusability guidelines. He mentioned that how much reusability resulted to improve product reliability and to reduce overall software development cost.

The problem of crosscutting which is produced during component development is elaborated (Clemente and Hernández, 2001). They solved this problem by the extension with Aspect oriented methodology. It was mentioned by an example that how new business rules resulted in the more adaptable and reusable components. According to them, this Aspect Component Based Software Engineering has been developed with success in the CORBA Component Model domain (Frakes and Kang, 2005).

The authors wrote a report for the validation of the component-based method (CBM) (Dolado, 2000) for software size estimation by the analysis of 46 projects. Then the complete process of this analysis and different techniques of analysis was mentioned. Relationship of LOC (Line of Code) and NOC (No. of Component) was carried out with suitable examples. Comparison of CBM and a Global Method (Mark II) (Symons, 1991) was also conducted (Dolado, 2000).

So far as this research is concerned, we have clearly analyzed that there is not a unique and single generalized method to find the SCE using CBSE. It may also be mentioned that SCE is a very crucial task. COCOMO-I&II are working in market to find SCE and different variants of the COCOMO-II as COGOMO and COCOTS etc are also in practice. The literature review reveals that there have been quite a number of limitations of these and some other existing cost estimation techniques that are summarized in Table 1.

**Problem statement:** There is neither a specialized model nor any generalized way to find out the SCE of an object-oriented software projects using CBSE. Hence there is a room for not only finding the significant parameters/variables for component-based cost estimation but also the level of their significance for the said cost estimation method.

**The proposed solution:** The proposed solution is to search the parameters/variables involved in SEC using CBSE, directly or indirectly. For this purpose, the authors searched the different parameters/variables for SCE of object-oriented projects using CBSE. These parameters/variables are searched through the literature review of the Int. journal papers of the area of SCE in object oriented environment using CBSE. These research papers are given in the reference section.

After searching the parameters/variables, a comprehensive survey is conducted to validate the proposed parameters/variables and to give the significance level of every parameter/variable to solve the problem.

Factor and Community analysis, statistical methods, is used by using the Special Package for the Social Sciences (SPSS) software, to evaluate the outcomes of the survey. This analysis will ultimately lead to find the results of the proposed solution (see Table 2 & Table 3 in section V), to provide the significance level of all the searched parameters/variables.

It may be mentioned here that there is not any research paper that can provide all the parameters/variables searched in our research work.

**Results of the research:** Following section is the detail explanation of the evaluation of the proposed solution (see section IV) using factor analysis with the help of SPSS software.

**Factor Analysis:** Factor analysis was used to reduce the leveling of parameters/variables and their significance for the SCE using CBSE. The principal component method was used to extract factors that have eigen values greater than none and the varimax rotation was used to make the factors more interpretable.

**Communalities Analysis:** To observe the importance of investigated variables/parameters, communality analysis was performed. Initial and extracted communalities values were reported in Table 2. Following is the definition of all the parameters/variables that are used in Table 2.

Definition of the Parameters/Variables Used In Table 2

| | | | |
|---|---|---|---|
| TFNC | ▶ Time consumed to find new component | ERCSS | ▶ Effort required for the component standardization in standalone environment |
| TCI | ▶ Time consumed to implement the components | ECSD | ▶ Effort needed for the component standardization in distributive environment |
| LCC | ▶ Level of complexity of the component | LSIC | ▶ Level of standardization of the interface of a component |
| DCI | ▶ Degree of component integration | LSFC | ▶ Level of standardization of functionality of the component |
| PIC: | ▶ Percentage of inherited component | QC | ▶ Quality of the component |
| FCUR | ▶ Frequency of changes in user requirements | LCCP | ▶ Level of customization of the Component(s) in project |

**Table 1: General Limitations of SCE Techniques that affects the overall cost of software project directly or indirectly**

| SCE using Algorithmic Methods | Limitations |
|---|---|
| LOC (Verner and Tate, 1992; Keung *et al.*, 2004; Albrecht and Gaftney, 1983) | 1. Post SCE<br>2. Language dependent<br>3. It may include substantial dead code and blanks.<br>4. Not properly working for visual languages<br>5. Difficult to count logical statements<br>6. Not widely automated.<br>7. Inconsistent for FP conversion |
| SLOC (Jones, 2006; Albrecht and Gaftney, 1983) | 1. No clear relationship between SLOC and the end product<br>2. Lack of normalized relation between programs because when language changes then SLOC also changes.<br>3. Programmer don't like SLOC<br>4. Size of software project affected by language. |
| SLIM (Putnam and Myers,1992; Putnam, 1978) | 1. Not appropriate for small projects<br>2. Its estimates are quite sensitive to the technology factor.<br>3. Not transparent |
| Function Point (Jones, 2006; Albrecht, 1979; Symons, 1988) | 1. Not suitable for scientific applications and real time control complex applications<br>2. Requirement of in-depth knowledge of standards for accurate counting.<br>3. Existence of some non standardized variations<br>4. Historical data does not exist like SLOC<br>5. Every now and then, backfiring from SLOC can be inaccurate and misleading<br>6. Subjective counting (screens, reports etc)<br>7. Hard to automate.<br>8. Ignores quality of output<br>9. Time consuming |
| Object-oriented Metrics (Jones, 2006) | 1. Does not support studies out side of object-oriented paradigm.<br>2. Does not deal with full life cycle issues<br>3. Difficult to itemize<br>4. No conversions rules to LOC and function point<br>5. Not supported by software estimating tools |
| COCOMO-I (Merlo-Schett, 2002) | 1. It is very difficult to accurately estimate Thousands Lines of Delivered Source Instructions (KDSI) early in the project when most estimates'-effort estimates are required.<br>2. Tremendously susceptible to misclassification of the development mode.<br>3. Achievement largely depends upon changing the model to the needs of an organization and availability of historical data that not available.<br>4. Project cost is being poorly estimated. |
| COCOMO-II (Mahmood *et al.*, 2005) | 1. It is still mainly based on water fall model<br>2. Most extensions of this model are still experimental and have not been fully used till now<br>3. Duration calculation of small projects is difficult |

This research work shows the main independent parameters/variables that have considerable effect on total cost of the software project using CBSE.

**Table 2: Communalities Analysis**

| Variable | Question asked | Initial | Extract |
|---|---|---|---|
| TFNC | Time consumed to find new component in the project. | 1.000 | .757 |
| TCI | Time consumed implement component in the project. | 1.000 | .695 |
| LCC | The level of complexity of the component in the project. | 1.000 | .378 |
| DCI | The degree of integration of the component with other component(s) in the project was. | 1.000 | .735 |
| PIC | The percentage of inherited component(s) in the project was. | 1.000 | .685 |
| FCUR | The frequency of changes in user requirements in the project was. | 1.000 | .725 |
| ERCSS | The effort required for the standardization of the component(s) to make them usable in standalone environment in the project was. | 1.000 | .724 |
| ECSD | The effort required for the standardization of the component(s) to make them usable in distributive environment in the project was. | 1.000 | .688 |
| LSIC | The level of standardization of interface of a component in the project was. | 1.000 | .505 |
| LSFC | The level of standardization of functionality of the component in the project was. | 1.000 | .636 |
| QC | The quality of the component in the project was. | 1.000 | .586 |
| LCCP | The level of customization of the Component(s) in the project was. | 1.000 | .597 |

It may be mentioned that most of the existing Algorithmic Cost Methods (Models) (Attarzadeh and Ow, 2010) given by different scientist/researchers are discussed in Related Work section. Since this paper gives an approach which is new and unique of its own type (To find the parameters/variables involved in SCE using CBSE and their significance levels). It means that there is not a need to compare the approach given by the authors of this paper with the existing Algorithmic Cost Methods (Attarzadeh and Ow, 2010) of software estimation discussed in related work.

Table 2 shows that the highest variation 76% is explained by time consumed to find new components TFNC, which suggested that TFNC is most significant variable for the ranking of variables or parameter for SCE of the software project using CBSE. It is observed that statistically least important variable is LCC among the other parameters/variables because the variation explained by LCC is 37.8 percent.

**Eigen Value Analysis:** Five factors or variables/parameters were extracted from the ranking of parameters/variables for SCE of the software project using CBSE. These are comprised of Standardization of components either in standalone environment or distributive environment (SC), Changing user requirements along with all the quality attributes of components (CURQA), Time consumed for detailed analysis of components (TDAC), Customization of newly developed or inherited components having efficient interfacing (CNIEI) and time consumed for the optimized implementation of the component (TOIC). The five factors account for 64.257 percent variation against 100 percent variation with twelve items. Percentage of variation explained by each variable with Cumulative percentage of variation is given in Table 3.

**Table 3: Percentage of Variance Explained By Each Factor**

| Component (Factor) | Initial Eign Values | | | Extraction Sums of Squared Loadings | | | Rotation Sums of Squared Loadings | | |
|---|---|---|---|---|---|---|---|---|---|
| | Total | % age of Var | Cum. % Var | Total | %age of Var | Cum.% Var | Total | %age of var | Cum. % |
| SC (F1) | 2.132 | 17.770 | 17.770 | 2.132 | 17.770 | 17.770 | 1.838 | 15.318 | 15.318 |
| CURQA (F2) | 1.702 | 14.179 | 31.950 | 1.702 | 14.179 | 31.950 | 1.578 | 13.151 | 28.468 |
| | 1.509 | 12.572 | 44.522 | 1.509 | 12.572 | 44.522 | 1.452 | 12.097 | 40.566 |
| TDAC (F3) | 1.333 | 11.105 | 55.627 | 1.333 | 11.105 | 55.627 | 1.425 | 11.878 | 52.443 |
| CNIEI (F4) | 1.036 | 8.630 | 64.257 | 1.036 | 8.630 | 64.257 | 1.418 | 11.813 | 64.257 |
| TOIC (F5) | | | | | | | | | |

Table 3 shows that existence of F1(Factor 1) explained 15.318 percent of total variation, therefore it is most highly contributing factor and continuous improvement for ranking for SCE using CBSE F5 (Factor 5) explained only 11.813 percent variation of the total variation, which is least contributing factor in five selected factors. Logical construct for total variance from twelve parameters/variables is given in Table 4.

**Table 4: Logical Construct for Total Variance from Twelve Variables/Parameters**

| Factor | No. of Variables | Cum. % |
|---|---|---|
| SC (F-1) | ERCSS, ECSD | 17.770 |
| CURQA (F-2) | DCI, FCUR, QC | 14.179 |
| (TDAC) F-3 | TFNC, LCC | 12.572 |
| CNIEI (F-4) | PIC, LSIC, LCCP | 11.105 |
| TOIC (F-5) | TCI, LSFC | 8.630 |

**Validity of research:** Survey methodology is used to validate the estimated software cost by visiting many National and Int. software companies namely NextBridge, Naseeb, Innovative, ITS, Techlogix, TRG, Smartix Sol., Title Development., Sigma Tech, IT Deptt. Govt. of Punjab, Vision Marketing & Mgt, Visionary Computer Sol., ESP Interactive sol., Trisoft, Warid Telecom, Corvit Systems, Technologies System Consulting, Systems, Secure-IT, Netsol Technologies, Trisoft & Mobilink Generation Systems. It is to be pointed out that the project analyst, project managers, team leads and senior developers evaluated our questionnaire. The suitable tact's and lessons in (Ji *et al.*, 2008) helped very much for the conduction of survey. Following are the assumptions that were made at the beginning of survey:

1. Project Managers or System Analyst are the key professionals for the evaluation of our survey.
2. All the software developers have more than 3 years of software development experience.
3. Software Company should me*e*t all the hardware and software requirements (e.g. Int. Standards Organization (ISO), Jovanovic and Shoemaker, (1997))
4. Software company has detailed know how about standards and quality of software (Jovanovic and Shoemaker, (1997). e.g. NetSol (a largest software company) has achieved CMMI level 5 certification.
5. The software company should have Int. clients as well as local clients.

**Conclusion:** As mentioned in Table 2 that eleven parameters/variables out of twelve parameters/variables are more than 50% momentous for the significant level of parameters/variables for SCE of the software project using CBSE. This implies that the core objective of this research paper has been achieved significantly. It may also be noted that the communality percentage of variance of the five factors calculated in Table 3 is 64.257.Hence the statistical analysis proves 64.257% as the accuracy or success level of this research. The main contribution of this research is to search the parameters/variables that have a significant role for the SCE using CBSE and not only find the level of significance of all the parameters/variables but also find the highest and least significant parameter/variable and overall significant effect of all the parameters/variables to find the SCE using CBSE. This research may act as an initial step towards SCE of complex systems using CBSE. The beneficiaries of this research are the researchers of the CBSE and the software industry.

**Limitations and future work:** As mentioned in section III, there is not any specialized method for the SCE of software projects using CBSE. Initially it was worrying that what would be the imitating step to estimate the software cost of project using components instead of classes, functions and attributes etc. A bold step was taken to read out the research papers that are discussed in section II. After literature review resulted in different parameters/variables that have vital role for the development of components. The research work done in this paper is to find out the significance of all the parameters/variables for the SCE of software systems. This research work does not apply to Expert Judgment and Machine Learning Methods. This paper does not calculate the exact estimated cost of software systems using CBSE and not give the specialized model or method but this is the first step towards the SCE of an object-oriented software systems using CBSE which will be accomplished through detailed afresh survey.

**Acknowledgements:** The authors would like to thank Dr. Tajammal Hussain and Mr. Muhammad Mohsin, assistant professors at department of mathematics, COMSATS Institute of Information technology (CIIT)Lahore Pakistan, for providing statistical analysis support. They are grateful to the CIIT Lahore Pakistan for providing valuable resources for the accomplishment of this research. They are also acknowledging the IT professionals and administration of software companies, at Lahore Pakistan, namely Next Bridge, Naseeb, Innovative, ITS, Techlogix, TRG, Smartix Sol., Title Development., Sigma Tech, IT Deptt. Govt. of Punjab, Vision Marketing & Mgt, Visionary Computer Sol., ESP Interactive sol., Trisoft, Warid Telecom, Corvit Systems, Technologies System Consulting, Systems, Secure-IT, Netsol Technologies, Trisoft & Mobilink Generation Systems whose kind cooperation played a vital role in the conduction of survey.